\documentclass[aps,pra,twocolumn,showpacs,floatfix]{revtex4-1}
\usepackage{graphicx}
\usepackage{hyperref}
\usepackage{color}
\usepackage{pgfplots}
\usepackage{latexsym,epsfig}
\usepackage{dcolumn}
\usepackage{amsthm,amsmath}

\begin{document}
\title{Dynamics of Resonant energy transfer in one-dimensional chain of Rydberg atoms}

\author{$^a$Maninder Kaur, $^a$Paramjit Kaur, $^b$B. K. Sahoo and $^a$Bindiya Arora}

\email{bindiya.phy@gndu.ac.in}

\affiliation{$^a$Department of Physics, Guru Nanak Dev University, Amritsar, Punjab-143005, India\\
$^b$Atomic, Molecular and Optical Physics Division, Physical Research Laboratory, Navrangpura, Ahmedabad-380009, India}

\date{Received date; Accepted date}

\begin{abstract}
We study resonant energy transfer in a one-dimensional chain of two to five atoms by analyzing time-dependent probabilities as function of their 
interatomic distances. The dynamics of the system are first investigated by including the nearest-neighbour interactions and then accounting for all 
next-neighbour interactions. We find that inclusion of nearest-neighbour interactions in the Hamiltonian for three atoms chain exhibits perdiocity during 
the energy transfer dynamics, however this behavior displays aperiodicity with the all-neighbour interactions. It shows for the equidistant chains
of four and five atoms the peaks are always irregular but regular peaks are retrieved when the inner atoms are placed closer than the atoms at both the ends. 
In this arrangement, the energy transfer swings between the atoms at both ends with very low probability of finding an atom at the center. This phenomenon 
resembles with quantum notion of Newton's cradle. We also find out the maximum distance up to which energy could be transferred within the typical 
lifetimes of the Rydberg states.
\end{abstract}

\maketitle

\section{Introduction}

Due to the peculiar nature, investigation of properties of Rydberg atoms have drawn attention in many branches of physics for decades. These atoms have one or more than one high-lying states with exaggerated responses to external electric and magnetic fields \cite{gallagher1994,dipoleblockade33}. Lifetimes of these highly-excited electronic states are very long compared to a
normal state in an atomic system and their wave functions can be described by the Bohr model of a hydrogen atom. Knowledge of lifetimes of Rydberg states can be useful to understand astrophysical
observations in the interstellar space~\cite{Gnedin} and new interactions of particle physics~\cite{Guise}. These atoms are also
suitable for investigating diamagnetic effects ~\cite{Neukammer}, plasma properties ~\cite{Vitrant}, quantum information technologies \cite{jak00}
etc. It has been of immense interest to fathom the interactions between the Rydberg atoms via collision processes owing to large sizes and dipole moments of these atoms \cite{PhysRevLett.47.405,
PhysRevLett.58.1324}. To investigate this, experiments were conducted initially at the room temperature. In such case the kinetic energies of the atoms are much larger than their interaction energies, so
the interactions were treated as perturbation to the thermal motion of the atoms \cite{PhysRevLett.47.405, PhysRevLett.58.1324}. However with the advent of laser cooling and trapping techniques of atoms, it has
enabled today to get deeper insights of the underlying interactions among the cold Rydberg atoms. These atoms provide ideal platforms to perform both the experiments and theoretical studies to investigate
possible exotic phenomena exhibiting within the systems. On the other hand, there have been attempts to invent quantum analogue of Newton's 
cradle, in which one observes nearly loss-less energy transfer between the first and last member of a hanging chain of metal balls that periodically swing 
with the central members staying still. A few of these studies include spin chain systems~\cite{pla_39, pla_40, pla_41} and optical lattice 
systems~\cite{pla_42}.

A particularly rewarding point in the detection of strong interactions between the Rydberg atoms is that the strength and characteristics of the interactions can be controlled experimentally. In
case of sufficiently strong interactions among the atoms, they can create entangled states with the surrounding atoms and behave in a collective fashion. Such states are of particular interest in
the context of proposed schemes for quantum information processing and for realizing fast quantum gates \cite{jak00}. The other important aspect of interparticle interactions is to study the phenomenon of resonant energy transfer (RET) where the state of one atom can greatly influence the other by exchanging their internal energies.
Interactions between the Rydberg atoms can be studied in the presence or absence of electric field~\cite{thadwalkerflannery,thadwalkerreihard,thadwalkersinger}.
In the region  of a high electric field, the atoms can acquire an induced dipole moment and thus, they can interact strongly like classical electric dipoles (interaction energy scales as $R^{-3}$, where $R$ is the interatomic distance between the atoms). Subjecting strong
electric field to atomic systems may help in creating strong interactions among the atoms, but it can unnecessarily bring in additional perturbative quantum effects. This can alter the original physical
properties of the atoms completely. Therefore, it is not advisable to use strong electric fields to study interactions of Rydberg atoms. In the absence of the electric field, interactions among the Rydberg
atoms can be enhanced by choosing  energetically degenerate transitions between the specific atomic states. In this situation two types of resonant processes can commonly occur; namely
the ``Fr\"oster" process and the ``migration'' process. In the Fr\"oster process initially two Rydberg atoms will be in the same state and they can undergo an energy
transfer process in such a way that one atom is transfered to the energetically lower-lying state while the other is transfered
to the higher-lying state. This process can be easily seen between two energetically degenerate Rydberg states.~\cite{for96,epjd}.
Natural occurrence of these events are rare but they can be achieved by tailoring the Rydberg atomic states by introducing small electric field \cite{dipoleblockade33}. For an appropriate field strength, two
transitions of a pair of atoms can be energetically degenerate owing to the Stark effect. In such case  the atoms can exchange energies, which is famously referred to as the RET
process \cite{epjd}. The ``migration'' process \cite{Comparat} on the other hand is a naturally existing resonant process in which a pair of
states of an atom automatically exchanges energy with identical pair of states of another atom. Though mediation by the external electric field is not required, but it necessitates to prepare the atoms in two different Rydberg states.
In this exactly RET reaction,
one atom gains exactly the same amount of energy that
the other one loses due to the non radiative dipole-dipole coupling. The periodic
transition of the system between the upper and lower electronic states of the atoms creates oscillating dipoles \cite{jd}. Since these interactions can govern naturally without the influence from the external electric field, it has several advantages to study them. This phenomena may also be applicable to
explain the cascading networks for complex energy transfer in the biological systems \cite{pair_int5,pair_int37,pair_int38,danielb,pair_int40,sho3,rit2}. Due to sufficiently long persistence of these resonant
behaviours in the Rydberg atoms, it is feasible to observe them experimentally and apply the energy transfer principle to various domains \cite{pair_int41}.

In this theoretical investigation, we intend to examine the atom-atom interactions that are enhanced by the migration processes for RET in a one-dimensional (1-D) chain of Rydberg atoms to
study their dynamical behavior. To gain some understanding about the collective behavior of these processes, we explore the effects of all-neighbour interactions among atoms in the 1-D chain and follow their dynamics by scanning through the interatomic distances between them.  We then extend the investigations further by including two more atoms in the chain to find out changes in the results. In these analyses, we have come up with few interesting observations
by varying interatomic distances, e.g. placing inner atoms closer than the atoms located at both the ends of a four and five atoms chain. 
It is observed that with proper choice of distances between the interior atoms of the chain, the system is capable of transferring the energy 
from the first atom to the last one by absorbing a negligible amount of energy. Such systems are highly desired due to their potential applications in 
building up loss-less quantum channels that can be used to connect quantum memories~\cite{pla}

The paper is organized as follows: In Sec. \ref{theory}, theoretical model adopted to study the interactions between two Rydberg atoms and the induced dipole-dipole force mediated by the migration processes
are presented. The numerical results for chains with different number of atoms are discussed in Sec. \ref{results}, followed by Conclusion in Sec. ~\ref{conclusion}.

\section{Theoretical Model }~\label{theory}

 \begin{figure}[t!]
 \includegraphics[width=8.5cm,height=9.5cm]{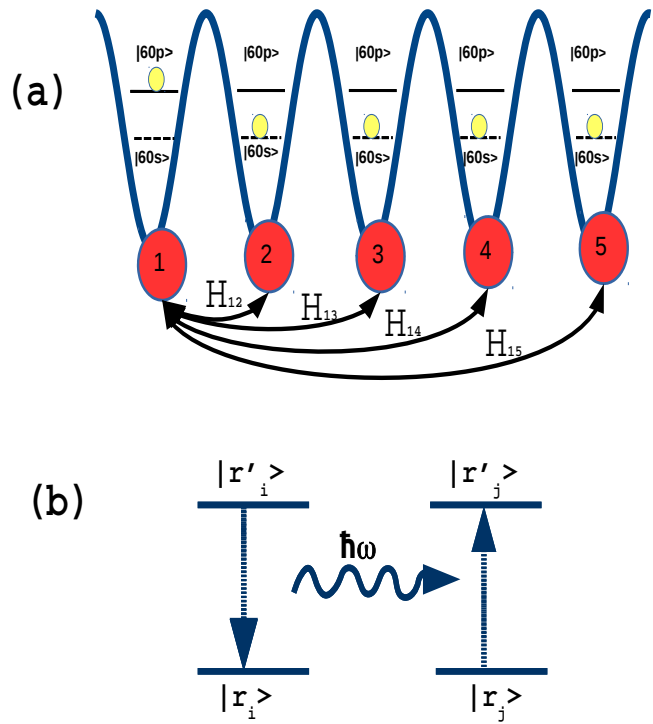}
  \caption{(Color online) Schematics of (a) one-dimensional chain of five atoms with one atom per lattice and (b) a typical migration resonant reaction
 $|r'_i\rangle+|r_j\rangle \leftrightarrow |r_i\rangle +|r'_j\rangle$ process from site $i$ to site $j$.}
 \label{fig-1n}
\end{figure}

A chain of Rydberg atoms with one atom per lattice site, as shown in Fig. \ref{fig-1n}(a), is undertaken in the present study. In the actual calculations, these Rydberg states are identified by their lifetimes and
dipole moments. We choose a hard-core lattice with two-level atom having lower state $|r_i \rangle$ of energy $E_{r_i}=\hbar \omega_{r_i}$ and  upper state $|r'_i \rangle$ of energy $E_{r'_i}=\hbar \omega_{r'_i}$
on the $i^{th}$ site , as can be seen in Fig. \ref{fig-1n}(b). The general Hamiltonian of such a system with $n$ lattice sites is given by
\begin{eqnarray}
H_{\rm{A}} &=&\hbar\sum_{i=1}^{n}\left [ \omega_{r_i}\vert r_i\rangle\langle r_i\vert +\omega_{r'_i}\vert r'_i\rangle\langle r'_i\vert\right ] .
\end{eqnarray}
In the matrix representation, choosing the kets as
\begin{eqnarray}
|r_i \rangle \equiv \left ( {\begin{array}{c}0 \\ 1 \\ \end{array} } \right )_i  \ \ \ \ \text{and} \ \ \ \ |r'_i \rangle \equiv \left ( {\begin{array}{c}1 \\ 0 \\ \end{array} } \right )_i,
\end{eqnarray}
we can rewrite the above Hamiltonian as
\begin{eqnarray}
 H_{\rm{A}} &=&\hbar\sum_{i=1}^{n}\left [ \omega_{r_i}\sigma_i^-\sigma_i^+ + \omega_{r'_i}\sigma_i^+ \sigma_i^-\right ] ,
  \label{eqhm}
 \end{eqnarray}
such that $\sigma_i^-$ and $\sigma_i^+$ act as the lowering and raising state operators with the forms
\begin{eqnarray}
\sigma_i^- \equiv \left ( {\begin{array}{cc}0 & 0 \\ 1 & 0 \\ \end{array} } \right )_i \ \ \ \ \text{and} \ \ \ \ \sigma_i^+ \equiv \left( {\begin{array}{cc}0 & 1 \\ 0 & 0 \\ \end{array} } \right )_i,
\end{eqnarray}
respectively. This follows $\sigma_i^- |r_i\rangle =0$,  $\sigma_i^+ |r_i\rangle =|r'_i\rangle$, $\sigma_i^- |r'_i\rangle =|r_i\rangle$, and $\sigma_i^+ |r'_i\rangle =0$.

We intend to investigate energy transfer through the migration process described by $\vert r'_i \rangle+\vert r_j \rangle\leftrightarrow \vert r_i \rangle+\vert r'_j\rangle$.  
To drive the process, we assume the first atom is in the excited state while the rest are in the lower level by defining the initial wave function of the composite system as
\begin{equation}
\vert \psi(0) \rangle=   \left ( {\begin{array}{c}1 \\ 0 \\ \end{array} } \right )_1\otimes \left ( {\begin{array}{c}0 \\ 1 \\ \end{array} } \right)_2\otimes\cdots\otimes\left( {\begin{array}{c}0 \\ 1 \\ \end{array} } \right)_n ,
\end{equation}
corresponding to the density operator
\begin{equation}
\rho(0)=\vert \psi(0) \rangle\langle \psi(0) \vert .
\end{equation}

An atom at the $i^{th}$ site in the upper level at a given time, when undergoes a transition, $|r'_i\rangle \rightarrow |r_i \rangle$, produces  an oscillating electric dipole moment by emitting a photon of energy $\hbar \omega=\hbar \omega_{r'_i} - \hbar \omega_{r_i}$. Because of this phenomena, the atom
present at the $j^{th}$ site can experience an induced electric field. The interaction Hamiltonian describing this process can be expressed as \cite{jd}
 \begin{equation}
 H_{\rm{int}}=\sum_{i<j}^{n}\left( H_{ij}^{r_i'r_i;r_jr_j'}+H_{ij}^{r_ir_i';r_j'r_j}\right),
 \end{equation}
where
\begin{eqnarray}
H_{ij}^{r_i'r_i;r_jr_j'} &=& \frac{\langle r'_i \vert \vec{\mu}_i \vert r_i \rangle \langle r_j \vert \vec{\mu}_j \vert r_j' \rangle }{4\pi\epsilon_0}\nonumber\\
&&\left[(1-3\cos^2\theta)\left(\frac{\cos(k R_{ij})}{R_{ij}^3}+\frac{k\sin(k R_{ij})}{R_{ij}^2}\right)\right.\nonumber\\
&&
-\left.\frac{k^2\cos(k R_{ij})\sin^2\theta}{R_{ij}}\right]
\left.( \sigma_i^-\sigma_j^+)\right. ,~\label{ih1}
\end{eqnarray}
and similarly for the Hamiltonian $H_{ij}^{r_ir_i';r_j'r_j}$. In this expression, $\theta$ is the angle between the radial distance $R_{ij}$ and induced electric field among the atoms,
$k=\omega /c$ is the wave number of the emitted photon, and the combined operators $\sigma_i^-\sigma_j^+$ indicate that the atom at the lattice site $i$ transits to the ground state and the atom at lattice site $j$  transfers to the excited
state when they act on the wave function of the system. It is also obvious from the above expression that it is imperative to have large transition dipole moments of the Rydberg atoms for strong interatomic interactions.

Now starting with $t_1=0$ for the density operator $\rho(0)$, the density operator of the system at time $t_n$ can be given as
\begin{eqnarray}
\rho(t_n) &=& U(t_n) \rho(t_{n-1}) U^{\dag}(t_n), \label{dm1}
\end{eqnarray}
where $U(t)=\exp \frac{-\iota H t}{\hbar}$ with the total Hamiltonian $H=H_{\rm{A}}+H_{\rm{int}}$. This means for the time-interval $\Delta t=t_n - t_{n-1}$, the
rate of change of density operator is given by
\begin{eqnarray}
\frac{d\rho(\Delta t)}{\Delta t} &=& \frac{dU(\Delta t)}{\Delta t} \rho(t_{n-1}) U^{\dag}(\Delta t) \nonumber \\ && + U(\Delta t) \rho(t_{n-1})\frac{dU^{\dag}(\Delta t)}{\Delta t}.
\end{eqnarray}
From this, we can obtain the Liouville's equation-of-motion \citep{bookblum} as
\begin{eqnarray}
\iota \hbar\frac{d\rho(t)}{dt} &=& [H, \rho(t)]\label{von}.
\end{eqnarray}
It, thus, implies that both Eqs. (\ref{dm1}) and (\ref{von}) are equivalent. We use Eq. (\ref{dm1}) for convenience to find out evolution of density of the system
over the time. It is also worth mentioning here that we have not taken into account the decay rate of the excited state in the above equation.

\begin{figure}[t!]
 \includegraphics[width=8.5cm,height=7.0cm]{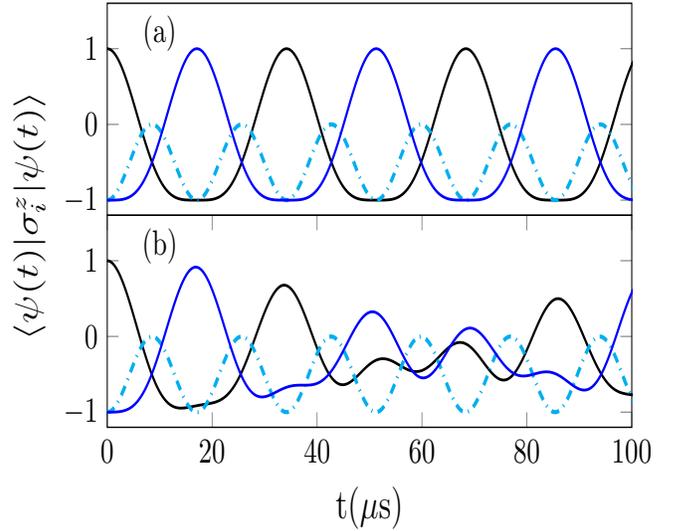}
  \caption{(Color online) Time evolution of $\langle \psi(t) \vert \sigma_i^z \vert \psi(t) \rangle$ for the chain of three atoms with interatomic distance $R_{i,i+1} = 95 \ \mu$m. Results including (a) only
the nearest-neighbour interactions and (b) all-neighbour interactions are shown. Here black solid, blue solid and cyan dash-dotted curves are for the first ($i=1$), second ($i=2$) and third ($i=3$) atom, respectively.}
 \label{fig-21}
\end{figure}

From the evolution of $\rho(t_n)$ over a time period, we can get signature of energy propagation from one site to other in the lattice during the migration process.
If the process is truly migration type then it can be quantified by finding out the chances of an atom being in the $|r_i\rangle $ and $|r'_i\rangle$ on the site $i$
at a given time. To identify this process quantitatively, we define the atomic inversion operator as
\begin{eqnarray}
\sigma_i^z &=& \vert r'_i \rangle \langle r'_i\vert-\vert r_i \rangle\langle r_i\vert\nonumber\\
          &=& \sigma_i^+ \sigma_i^- -\sigma_i^-\sigma_i^+.
\end{eqnarray}
The expectation value of the above operator, $\langle \psi(t) \vert \sigma_i^z \vert \psi(t) \rangle$, can be interpreted as the difference between the probability of occupation for the  $\vert r_i \rangle$ state to the $\vert r'_i \rangle$ state on the site $i$ at time $t$ as follows. Let us express the wave function of the atom on the site $i$ as a linear
combination of both the possible states at a given time $t$ as $|\phi_i(t) \rangle = c_{r_i}(t) \vert r_i \rangle + c_{r'_i}(t) \vert r'_i \rangle$, where
$c_{r_i}(t)$ and $c_{r'_i}(t)$ are the probability coefficients of an atom being in the $\vert r_i \rangle$ and $ \vert r'_i \rangle$ states,  respectively, such that
\begin{eqnarray}
|c_{r_i}(t)|^2 + |c_{r'_i}(t)|^2=1 .
\end{eqnarray}
We can now write
\begin{equation}
\vert \psi(t) \rangle= |\phi_1(t) \rangle \otimes |\phi_2(t) \rangle \otimes \cdots \otimes |\phi_n(t) \rangle .
\end{equation}
Using this wave function, it yields
\begin{equation}
\langle \psi(t) \vert \sigma_i^z \vert \psi(t) \rangle= Tr[\sigma_i^z \rho(t)] =\vert c_{r'_i} \vert^2-\vert c_{r_i} \vert^2 .
\end{equation}
Obviously, its value can vary in between $-1$ to $1$ in the migration process. In other words, at time $t$ if the atom on the site $i$ is in the $\vert r_i \rangle$
state then $\langle \psi(t) \vert \sigma_i^z \vert \psi(t) \rangle=-1$, and if it is in the $\vert r'_i \rangle$ state then $\langle \psi(t) \vert \sigma_i^z
\vert \psi(t) \rangle=1$.

\begin{figure}[t!]
 \includegraphics[width=8.5cm,height=8.5cm]{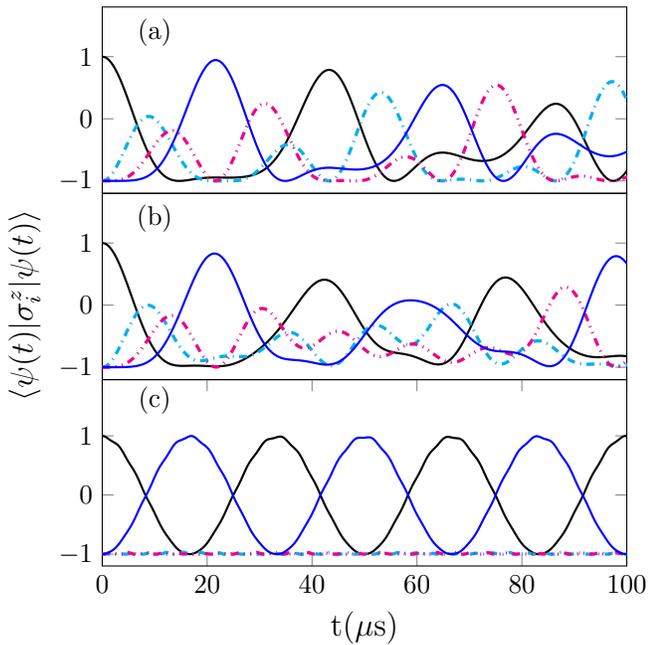}
  \caption{(Color online) Time evolution of $\langle \psi(t) \vert \sigma_i^z \vert \psi(t) \rangle$ for the chain of four atoms. Results (a) with only the nearest-neighbour interactions and equal distance
$R_{i,i+1}= 95 \ \mu$m;(b) considering all-neighbour interactions in the atoms placed at equal distance $R_{i,i+1}= 95 \ \mu$m; and (c) considering the nearest-neighbour interactions between non-equidistant atoms with interatomic distances $R_{12} = 42 \ \mu$m, $R_{23} = 16.8 \ \mu$m and $R_{34} = 42 \ \mu$m are shown. Here black, cyan dash-dotted, magenta dash double-dotted and blue solid curves 
are for the first ($i=1$), second ($i=2$), third ($i=3$) and forth ($i=4$) atoms, respectively.}
 \label{fig-4}
\end{figure}

\begin{figure}[t!]
 \includegraphics[width=8.5cm,height=8.5cm]{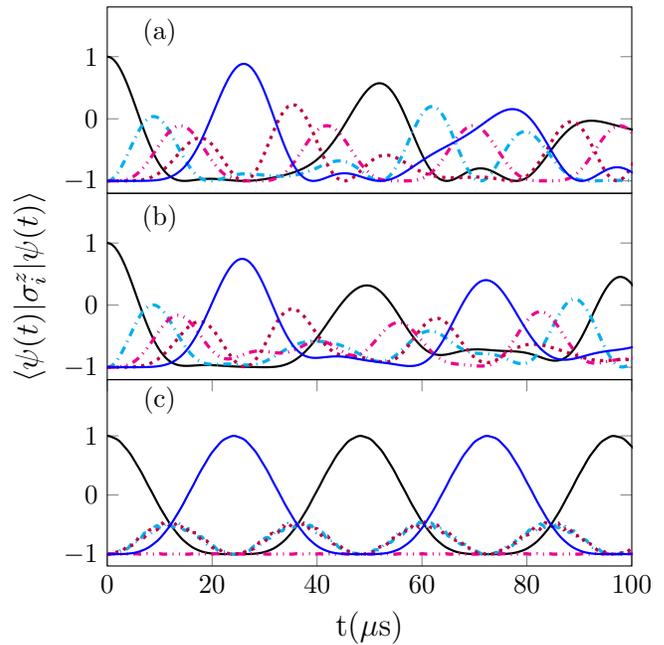}
  \caption{(Color online) Time evolution of $\langle \psi(t) \vert \sigma_i^z \vert \psi(t) \rangle$ for the chain of five atoms. They are shown for the cases (a)
considering  only the nearest-neighbour interactions among the atoms with equal distance $R_{i,i+1}= 95 \ \mu$m; (b)with  all-neighbour interactions in the atoms placed at equal distance with $R_{i,i+1}= 95 \ \mu$m; and (c) with the nearest-neighbour interactions for non- equidistant  chain with interatomic distances
$R_{12} = 95 \ \mu$m, $R_{23} = 38 \ \mu$m, $R_{34} = 38 \ \mu$m, $R_{45} = 95 \ \mu$m, respectively. The black,
cyan dash-dotted, magenta dash-double-dotted, blue solid and purple dotted curves represent the  first ($i=1$), second ($i=2$), third ($i=3$), fourth ($i=4$) and fifth  ($i=5$) atoms,
respectively.}
 \label{fig-5}
\end{figure}

\section{Results and discussions}~\label{results}

\subsection{Dynamics of atomic inversion}

To understand RET in the migration process, we have evaluated the dynamics of states via the temporal evolution of the $\langle \psi(t) \vert \sigma_i^z \vert \psi(t) \rangle$ values.  We would like to study the transition probability of the 
$\vert 60s\rangle + \vert 60p\rangle \leftrightarrow \vert 60p\rangle + \vert 60s \rangle$ transition in the migration process. We have estimated the lifetimes of the $\vert 60s \rangle$ and $\vert 60p \rangle$ Rydberg states 
of the Rb atom as about 206 μs and 564 μs, respectively, using the 
fitting formula of the quantum defect theory as \cite{gounand}
\begin{equation}
\tau_nl = \tau_l (n-\delta_l)^{\epsilon_l}.
\end{equation}

In this expression, $\tau_l$, $\delta_l$ and $\epsilon_l$ are the suitable 
parameters for the respective states that are taken from \cite{branden}.
In our analysis, we restrict evolution of density of the wave function in the above composite system up to 100 $\mu$s. This is sufficiently less 
than the lifetimes of the above states. We present the transient behaviour of RET in two different regimes. We consider only the nearest neighbouring interactions in the first case  whereas in 
the second case, we allow all-neighbouring interactions in the interaction Hamiltonian. This includes all possible pair interactions among the atoms. We present results below separately for 
a chain of atoms with three, four and five atoms in the lattice.

\begin{figure}[t!]
 \includegraphics[width=8.5cm,height=5.0cm]{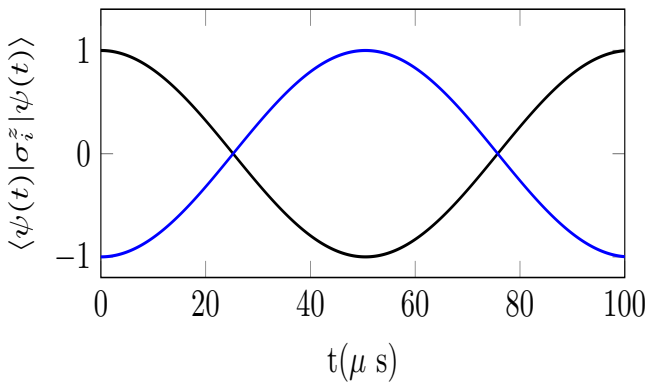}
  \caption{(Color online) Time evolution of $\langle \psi(t) \vert \sigma_i^z \vert \psi(t) \rangle$ for two atoms with interatomic distance $R_{i,i+1}=153 \ \mu$m. Here, 
the black and blue solid curves are for the first ($i=1$) and second ($i=2$) atoms, respectively.}
\label{fig-8}
\end{figure}

At the very outset, we analyze evolution of $\langle \psi(t) \vert \sigma_i^z \vert \psi(t) \rangle$ for an equidistant chain of three individually addressable Rydberg atoms with interatomic 
distance $R_{i,i+1}=  95\ \mu$m between two consecutive atoms. At time $t = 0 \mu$s, it is assumed that the first atom is in the excited Rydberg state, $|60p\rangle$, whereas the second and third 
atoms are in the Rydberg state, $|60s\rangle$.  Once the first atom is transfered from the $|60p\rangle$ state to the $|60s\rangle$ state, a photon is released which 
 excites the second atom to $|60p\rangle$. Subsequently excitation hopping of the photon takes place between the second and third atoms. The Fig. \ref{fig-21}(a) shows the dynamics for 
temporal evolution of $\langle \psi(t) \vert \sigma_i^z \vert \psi(t) \rangle$ including only the nearest neighbour interactions in $H_{\rm int}$. As can be seen, states of the atoms oscillate 
back and forth between $-1$ and 1 for the first and third atoms with a fixed frequency over a period of time. This frequency of oscillation is proportional to the magnitude of the induced 
electric field felt by an atom due to the presence of other atoms in the chain. While for the atom in the middle, the probability oscillates twice as fast between $-1$ and 0. This observation is a 
clear indication of the presence of the electric field two times stronger at the center of the chain as compared to the electric fields at the extreme ends.
When all-neighbour interactions are allowed through the $H_{12}$, $H_{23}$ and $H_{13}$ terms in $H_{\rm int}$, we observe aperiodicity  in the evolution of $\langle \psi(t) \vert \sigma_i^z \vert \psi(t) \rangle$ 
as shown in Fig. \ref{fig-21}(b). This irregularity arises due to the competition between the nearest-neighbour and other next-neighbour interactions. Such patterns are in accordance with the experimental 
observations performed by Barredo {\it et al.} \cite{danielb}. It would be interesting to investigate roles of these interactions in the chains with four and five atoms.

\begin{figure}[t!]
 \includegraphics[width=8cm,height=7cm]{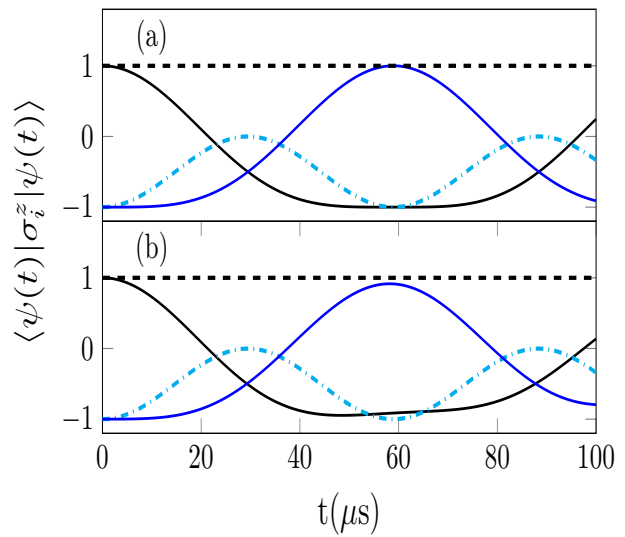}
  \caption{(Color online) Time evolution of $\langle \psi(t) \vert \sigma_i^z \vert \psi(t) \rangle$ for the chain of three atoms with interatomic distance $R_{i,i+1}=144 \ \mu$m. Results considering 
(a) only the nearest neighbour interactions and (b) all-neighbour interactions are shown. Here the black solid, cyan dash dotted and blue solid curves correspond to the first ($i=1$), second ($i=2$) 
and third ($i=3$) atoms, respectively.}
\label{fig-9}
\end{figure}

Now we extend the system to a four atom equidistant chain, with a distance $R_{i,i+1}=95\ \mu$m between two consecutive atoms. Fig. \ref{fig-4}(a) corresponds to the case where only the nearest 
neighbour interactions are retained in $H_{\rm int}$. One clearly observes from this figure that the dynamics of the probability of occupation of a state is aperiodic in contrast to the situation 
as was shown in Fig.~\ref{fig-21}(a). This indicates a competitive interplay between the
nearest neighbour interactions among the atoms. This again shows aperiodicity when all possible pair interactions among the atoms are introduced through $H_{\rm int}$ as shown in Fig.
\ref{fig-4}(b). However for an alternate arrangement of the atoms in the chain in which
the distance between first and second atom has been taken equal to the distance between third and forth atom, $R_{12} = R_{34}= x_o\ \mu$m(say) and the central atoms are placed such that $R_{23} = 0.4 \times x_o \ \mu$m the  distance values $R_{12} = R_{34}= 42 \ \mu$m and $R_{23} = 16.8 \ \mu$m, the periodicity is regained (see Fig. \ref{fig-4}(c)). 
Moreover in the non-equidistant arrangement of the atoms in a chain, the dynamics of the state at the site $i$ represented by $\langle \psi(t) \vert \sigma_i^z \vert \psi(t) \rangle$ oscillates periodically between $-1$ and $1$
for the first and last atoms in the lattice and with very small amplitudes for the oscillations of states are realized for the atoms placed in between this arrangement as shown in Fig. \ref{fig-4}(c) much like a classical Newton's cradle.
We have scanned through various value of $x_o$ starting from $ x_o=95 \ \mu$m upto a few micrometers and have observed a similar behaviour of excitation exchange. 

In continuation with above discussion, we present next the dynamics of $\langle \psi(t) \vert \sigma_i^z \vert \psi(t) \rangle$ in a chain of five atoms.  We show evolution of $\langle \psi(t) \vert \sigma_i^z \vert \psi(t) \rangle$
with the nearest neighbour and all pair interactions in Figs. \ref{fig-5}(a) and \ref{fig-5}(b), respectively. Then, 
we analyze these data for a non-equidistant chain of five atoms including only the nearest neighbour interactions and show them in \ref{fig-5}(c)).  The dynamics of $\langle \psi(t) \vert \sigma_i^z \vert \psi(t) \rangle$ are found 
to be similar to the case of the four atoms chain.
 	
\begin{figure}[t!]
\includegraphics[width=8.5cm,height=7cm]{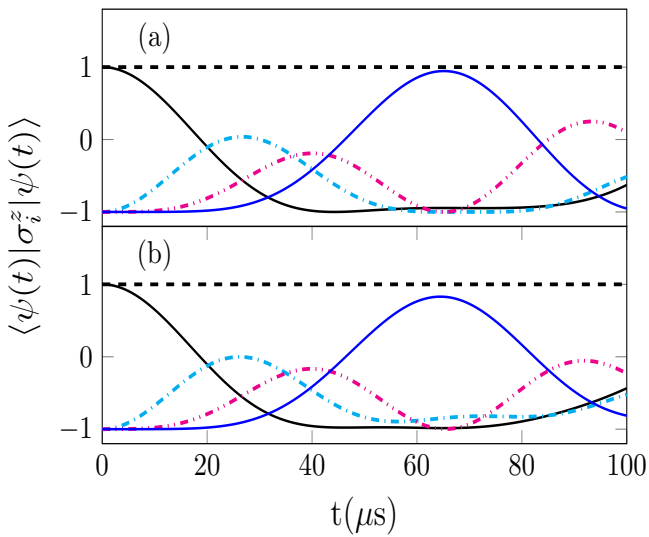}
  \caption{(Color online) Time evolution of $\langle \psi(t) \vert \sigma_i^z \vert \psi(t) \rangle$ for a chain of four atoms  with interatomic distance  $R_{i,i+1}=137 \ \mu$m. We show results
including (a) only the nearest neighbour interactions and (b) all-neighbour interactions. Here the black solid, cyan dash-dotted, magenta dash-double-dotted and blue solid curves correspond to
the first ($i=1$), second ($i=2$), third ($i=3$) and fourth ($i=4$) atoms respectively.}
 \label{fig-11}
 \end{figure}

\begin{figure}[t!]
 \includegraphics[width=8.5cm,height=7cm]{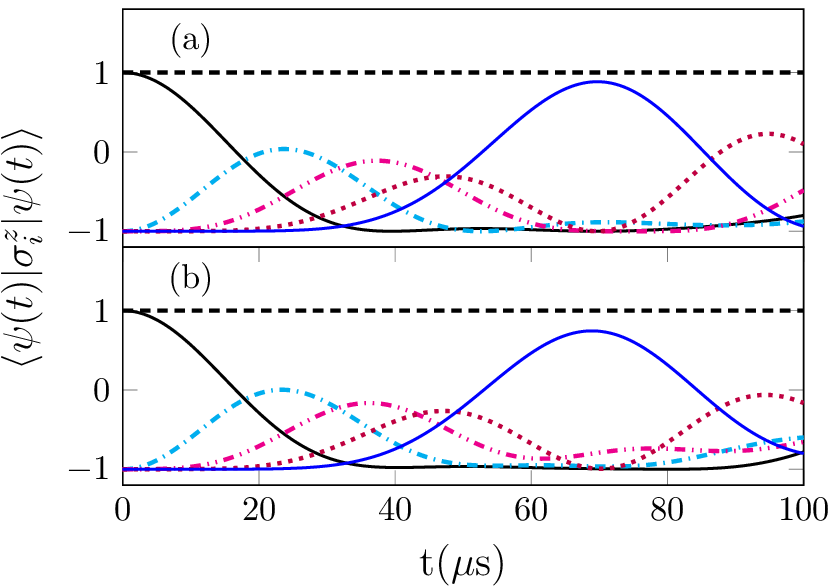}
  \caption{(Color online) Time evolution of $\langle \psi(t) \vert \sigma_i^z \vert \psi(t) \rangle$ for the chain of five atoms  withinteratomic distance $R_{i,i+1}=132 \ \mu$m. We have shown results
with (a) only the nearest neighbour interactions and (b) all-neighbour interactions. Here the black solid, cyan dash-dotted, magenta dash-double-dotted, blue solid and red dotted curves correspond to
the first ($i=1$), second ($i=2$), third ($i=3$), fourth ($i=4$) and fifth ($i=5$) atoms, repectively.}
 \label{fig-12}
 \end{figure}

\subsection{Energy transfer channel}

Here, we discuss the practical aspects of building up fast and efficient energy transfer channels using the cascade of atoms in one-dimensional chain.  It is obvious from Eq. (\ref{ih1}) that the dominant dipole interactions between the Rydberg atoms is proportional to $1/R_{ij}^3$. So, there must exist a maximum distance beyond which
the dipole interactions are not sufficient enough to enable propagation of energy. We want to find out an optimum distance such that excitation exchange between the first and the last atom in the chain occurs atleast once   within  100 $\mu$s. We find out the roles played by the 
nearest-neighbour and all-neighbour interactions in deciding this optimum distance and probability with which this excitation exchange of energy  can take place. Below  we present the RET  within the speculated time limit in the systems containing two to five  atoms in the chain.

In Fig.~\ref{fig-8}, we show the temporal evolution of $\langle \psi(t) \vert \sigma_i^z \vert \psi(t) \rangle$, which is directly related to the probability of energy transfer. We found that exchange of excitation between these atoms is possible via the migration process up to a maximum distance of $R_{i,i+1}= 153 \ \mu$m. 
Therefore, it looks like RET is not possible between the chosen Rydberg states of Rb atom beyond this distance.

Next, we model RET along a chain of three atoms. We show temporal evolution of $\langle \psi(t) \vert \sigma_i^z \vert \psi(t) \rangle$ of this system in Fig. \ref{fig-9}(a) by allowing only the nearest
interactions. We found that the maximum distance between the consecutive atoms for which the photon can be transferred atleast once  between the first and third  atom is $R_{i,i+1}= 144 \ \mu$m. Further investigation including all-neighbour interactions reveals that this optimum distance to transport a photon remains unchanged, but the maximum 
amplitude of $\langle \psi(t) \vert \sigma_i^z \vert \psi(t) \rangle$ reduces about 8\% as compared to the case of a system of three atoms governed only due to the nearest neighbour interactions 
as shown in Fig. \ref{fig-9}(b).

\begin{figure}[t!]
 \includegraphics[width=8.0cm,height=6cm]{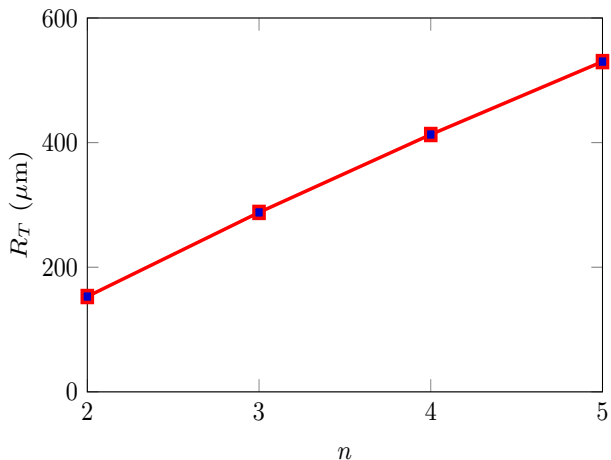}
  \caption{(Color online) The estimated maximum distance $R_T = (n-1)R_{i,i+1}$ for which the photon transfer takes place along the chain of a one-dimensional Rydberg atoms versus the number of atoms  $n$  in the chain.}
 \label{fig-13}
 \end{figure}

In Figs. \ref{fig-11}(a) and \ref{fig-12}(a),  we have shown time dependence of $\langle \psi(t) \vert \sigma_i^z \vert \psi(t) \rangle$ for the systems with four and five atoms, respectively. The maximum distance between two consecutive atoms which allow a photon transfer in a four atoms chain with the nearest
neighbour interactions is $R_{i,i+1} = 137~\mu$m, whereas it reduces to $132~\mu$ m for
a five atoms chain.  The value of $R_{i,i+1}$ did not change for the respective chain with the inclusion of all-neighbour interactions as shown in Figs. \ref{fig-11}(b) and \ref{fig-12}(b). Although, the maximum probability for photon transfer further reduced by 17 \% for a four atom chain and 
26\% for a five atom chain when all possible pairs of interactions were included. This signifies the fact that as the number of
atoms in the chain increases, probability of occurrence of RET through
the migration process decreases. 
From Fig.\ref{fig-13}, one can find out the maximum distance $R_T = (n-1)R_{i,i+1}$ up to which a photon can be
transferred from one end  to the other of an $n$ atoms chain. It can be estimated from this figure that a photon can be transferred up to
a maximum distance of about
$R_T=528 \ \mu$m in the migration process of a five atoms chain. 

\section{Concluding Remarks}~\label{conclusion}

We analyse the Rydberg-Rydberg interactions mediated by migration resonances in one-dimensional  equidistant as well as non-equidistant chains of two to five Rydberg atoms for their possible applications in building up the resonant energy transfer channels.  We have
explored the fundamental nature of the system dynamics considering only the nearest neighbour interactions and then considering all possible pairs of interactions. Our results on one hand show
robustness of the resonant energy transfer process within a range of parameters and on the other hand provide information on the limits of the interatomic distances after which energy exchange is not possible. Moreover, increase in the strengths of interactions among the atoms due to presence of more number of atoms affects the probability of the photon transfer. With the specific choices
of parameters corresponding to the $\vert 60s \rangle$ and $\vert 60p \rangle$ Rydberg states, we predict a maximum distance of $528$ $\mu$m for the energy transfer to occur along the
complete length of a five atoms chain. This is reasonable in the
atomic scale to be realized in a suitable quantum system. This may open up new perspectives in the application of resonant energy transfer process in constructing fast and efficient energy transfer 
channels.
 A non-equidistance chain of four and five  Rydberg atoms provides a system analogous to classical Newton's cradle in which the energy from the the first atom tunnels to the last atom through the channel constituted by the central atoms that practically do not absorb enegy.

\section*{Acknowledgements}

The authors would like to thank Prof. H. Ott for reading and making comments on the manuscript. The work of B.A. is supported by DST-SERB Grant No. EMR/2016/001228.

%

\begin{thebibliography}{32}%
\makeatletter
\providecommand \@ifxundefined [1]{%
 \@ifx{#1\undefined}
}%
\providecommand \@ifnum [1]{%
 \ifnum #1\expandafter \@firstoftwo
 \else \expandafter \@secondoftwo
 \fi
}%
\providecommand \@ifx [1]{%
 \ifx #1\expandafter \@firstoftwo
 \else \expandafter \@secondoftwo
 \fi
}%
\providecommand \natexlab [1]{#1}%
\providecommand \enquote  [1]{``#1''}%
\providecommand \bibnamefont  [1]{#1}%
\providecommand \bibfnamefont [1]{#1}%
\providecommand \citenamefont [1]{#1}%
\providecommand \href@noop [0]{\@secondoftwo}%
\providecommand \href [0]{\begingroup \@sanitize@url \@href}%
\providecommand \@href[1]{\@@startlink{#1}\@@href}%
\providecommand \@@href[1]{\endgroup#1\@@endlink}%
\providecommand \@sanitize@url [0]{\catcode `\\12\catcode `\$12\catcode
  `\&12\catcode `\#12\catcode `\^12\catcode `\_12\catcode `\%12\relax}%
\providecommand \@@startlink[1]{}%
\providecommand \@@endlink[0]{}%
\providecommand \url  [0]{\begingroup\@sanitize@url \@url }%
\providecommand \@url [1]{\endgroup\@href {#1}{\urlprefix }}%
\providecommand \urlprefix  [0]{URL }%
\providecommand \Eprint [0]{\href }%
\providecommand \doibase [0]{http://dx.doi.org/}%
\providecommand \selectlanguage [0]{\@gobble}%
\providecommand \bibinfo  [0]{\@secondoftwo}%
\providecommand \bibfield  [0]{\@secondoftwo}%
\providecommand \translation [1]{[#1]}%
\providecommand \BibitemOpen [0]{}%
\providecommand \bibitemStop [0]{}%
\providecommand \bibitemNoStop [0]{.\EOS\space}%
\providecommand \EOS [0]{\spacefactor3000\relax}%
\providecommand \BibitemShut  [1]{\csname bibitem#1\endcsname}%
\let\auto@bib@innerbib\@empty
\bibitem [{\citenamefont {Gallagher}(1994)}]{gallagher1994}%
  \BibitemOpen
  \bibfield  {author} {\bibinfo {author} {\bibfnamefont {T.~F.}\ \bibnamefont
  {Gallagher}},\ }\href {\doibase 10.1017/CBO9780511524530} {\emph {\bibinfo
  {title} {Rydberg Atoms}}},\ Cambridge Monographs on Atomic, Molecular and
  Chemical Physics\ (\bibinfo  {publisher} {Cambridge University Press},\
  \bibinfo {year} {1994})\BibitemShut {NoStop}%
\bibitem [{\citenamefont {Zimmerman}\ \emph {et~al.}(1979)\citenamefont
  {Zimmerman}, \citenamefont {Littman}, \citenamefont {Kash},\ and\
  \citenamefont {Kleppner}}]{dipoleblockade33}%
  \BibitemOpen
  \bibfield  {author} {\bibinfo {author} {\bibfnamefont {M.~L.}\ \bibnamefont
  {Zimmerman}}, \bibinfo {author} {\bibfnamefont {M.~G.}\ \bibnamefont
  {Littman}}, \bibinfo {author} {\bibfnamefont {M.~M.}\ \bibnamefont {Kash}}, \
  and\ \bibinfo {author} {\bibfnamefont {D.}~\bibnamefont {Kleppner}},\ }\href
  {\doibase 10.1103/PhysRevA.20.2251} {\bibfield  {journal} {\bibinfo
  {journal} {Phys. Rev. A}\ }\textbf {\bibinfo {volume} {20}},\ \bibinfo
  {pages} {2251} (\bibinfo {year} {1979})}\BibitemShut {NoStop}%
\bibitem [{\citenamefont {Gnedin}\ \emph {et~al.}(2009)\citenamefont {Gnedin},
  \citenamefont {Mihajlov}, \citenamefont {Ignjatović}, \citenamefont {Sakan},
  \citenamefont {Srećković}, \citenamefont {Zakharov}, \citenamefont
  {Bezuglov},\ and\ \citenamefont {Klycharev}}]{Gnedin}%
  \BibitemOpen
  \bibfield  {author} {\bibinfo {author} {\bibfnamefont {Y.}~\bibnamefont
  {Gnedin}}, \bibinfo {author} {\bibfnamefont {A.}~\bibnamefont {Mihajlov}},
  \bibinfo {author} {\bibfnamefont {L.}~\bibnamefont {Ignjatović}}, \bibinfo
  {author} {\bibfnamefont {N.}~\bibnamefont {Sakan}}, \bibinfo {author}
  {\bibfnamefont {V.}~\bibnamefont {Srećković}}, \bibinfo {author}
  {\bibfnamefont {M.}~\bibnamefont {Zakharov}}, \bibinfo {author}
  {\bibfnamefont {N.}~\bibnamefont {Bezuglov}}, \ and\ \bibinfo {author}
  {\bibfnamefont {A.}~\bibnamefont {Klycharev}},\ }\href {\doibase
  https://doi.org/10.1016/j.newar.2009.07.003} {\bibfield  {journal} {\bibinfo
  {journal} {New Astron. Rev.}\ }\textbf {\bibinfo {volume} {53}},\ \bibinfo
  {pages} {259 } (\bibinfo {year} {2009})},\ \bibinfo {note} {proceedings of
  the VII Serbian Conference on Spectral Line Shapes (VII SCSLSA) held in
  Zrenjanin, Serbia June 15th-19th 2009}\BibitemShut {NoStop}%
\bibitem [{\citenamefont {Guise}\ \emph {et~al.}(2014)\citenamefont {Guise},
  \citenamefont {Tan}, \citenamefont {Brewer}, \citenamefont {Fischer},\ and\
  \citenamefont {J\"onsson}}]{Guise}%
  \BibitemOpen
  \bibfield  {author} {\bibinfo {author} {\bibfnamefont {N.~D.}\ \bibnamefont
  {Guise}}, \bibinfo {author} {\bibfnamefont {J.~N.}\ \bibnamefont {Tan}},
  \bibinfo {author} {\bibfnamefont {S.~M.}\ \bibnamefont {Brewer}}, \bibinfo
  {author} {\bibfnamefont {C.~F.}\ \bibnamefont {Fischer}}, \ and\ \bibinfo
  {author} {\bibfnamefont {P.}~\bibnamefont {J\"onsson}},\ }\href {\doibase
  10.1103/PhysRevA.89.040502} {\bibfield  {journal} {\bibinfo  {journal} {Phys.
  Rev. A}\ }\textbf {\bibinfo {volume} {89}},\ \bibinfo {pages} {040502}
  (\bibinfo {year} {2014})}\BibitemShut {NoStop}%
\bibitem [{\citenamefont {Neukammer}\ \emph {et~al.}(1984)\citenamefont
  {Neukammer}, \citenamefont {Rinneberg},\ and\ \citenamefont
  {Majewski}}]{Neukammer}%
  \BibitemOpen
  \bibfield  {author} {\bibinfo {author} {\bibfnamefont {J.}~\bibnamefont
  {Neukammer}}, \bibinfo {author} {\bibfnamefont {H.}~\bibnamefont
  {Rinneberg}}, \ and\ \bibinfo {author} {\bibfnamefont {U.}~\bibnamefont
  {Majewski}},\ }\href {\doibase 10.1103/PhysRevA.30.1142} {\bibfield
  {journal} {\bibinfo  {journal} {Phys. Rev. A}\ }\textbf {\bibinfo {volume}
  {30}},\ \bibinfo {pages} {1142} (\bibinfo {year} {1984})}\BibitemShut
  {NoStop}%
\bibitem [{\citenamefont {Vitrant}\ \emph {et~al.}(1982)\citenamefont
  {Vitrant}, \citenamefont {Raimond}, \citenamefont {Gross},\ and\
  \citenamefont {Haroche}}]{Vitrant}%
  \BibitemOpen
  \bibfield  {author} {\bibinfo {author} {\bibfnamefont {G.}~\bibnamefont
  {Vitrant}}, \bibinfo {author} {\bibfnamefont {J.~M.}\ \bibnamefont
  {Raimond}}, \bibinfo {author} {\bibfnamefont {M.}~\bibnamefont {Gross}}, \
  and\ \bibinfo {author} {\bibfnamefont {S.}~\bibnamefont {Haroche}},\ }\href
  {http://stacks.iop.org/0022-3700/15/i=2/a=004} {\bibfield  {journal}
  {\bibinfo  {journal} {J. Phys. B. At. Mol. Opt. Phys.}\ }\textbf {\bibinfo
  {volume} {15}},\ \bibinfo {pages} {L49} (\bibinfo {year} {1982})}\BibitemShut
  {NoStop}%
\bibitem [{\citenamefont {Jaksch}\ \emph {et~al.}(2000)\citenamefont {Jaksch},
  \citenamefont {Cirac}, \citenamefont {Zoller}, \citenamefont {Rolston},
  \citenamefont {C\^ot\'e},\ and\ \citenamefont {Lukin}}]{jak00}%
  \BibitemOpen
  \bibfield  {author} {\bibinfo {author} {\bibfnamefont {D.}~\bibnamefont
  {Jaksch}}, \bibinfo {author} {\bibfnamefont {J.~I.}\ \bibnamefont {Cirac}},
  \bibinfo {author} {\bibfnamefont {P.}~\bibnamefont {Zoller}}, \bibinfo
  {author} {\bibfnamefont {S.~L.}\ \bibnamefont {Rolston}}, \bibinfo {author}
  {\bibfnamefont {R.}~\bibnamefont {C\^ot\'e}}, \ and\ \bibinfo {author}
  {\bibfnamefont {M.~D.}\ \bibnamefont {Lukin}},\ }\href {\doibase
  10.1103/PhysRevLett.85.2208} {\bibfield  {journal} {\bibinfo  {journal}
  {Phys. Rev. Lett.}\ }\textbf {\bibinfo {volume} {85}},\ \bibinfo {pages}
  {2208} (\bibinfo {year} {2000})}\BibitemShut {NoStop}%
\bibitem [{\citenamefont {Safinya}\ \emph {et~al.}(1981)\citenamefont
  {Safinya}, \citenamefont {Delpech}, \citenamefont {Gounand}, \citenamefont
  {Sandner},\ and\ \citenamefont {Gallagher}}]{PhysRevLett.47.405}%
  \BibitemOpen
  \bibfield  {author} {\bibinfo {author} {\bibfnamefont {K.~A.}\ \bibnamefont
  {Safinya}}, \bibinfo {author} {\bibfnamefont {J.~F.}\ \bibnamefont
  {Delpech}}, \bibinfo {author} {\bibfnamefont {F.}~\bibnamefont {Gounand}},
  \bibinfo {author} {\bibfnamefont {W.}~\bibnamefont {Sandner}}, \ and\
  \bibinfo {author} {\bibfnamefont {T.~F.}\ \bibnamefont {Gallagher}},\ }\href
  {\doibase 10.1103/PhysRevLett.47.405} {\bibfield  {journal} {\bibinfo
  {journal} {Phys. Rev. Lett.}\ }\textbf {\bibinfo {volume} {47}},\ \bibinfo
  {pages} {405} (\bibinfo {year} {1981})}\BibitemShut {NoStop}%
\bibitem [{\citenamefont {Stoneman}\ \emph {et~al.}(1987)\citenamefont
  {Stoneman}, \citenamefont {Adams},\ and\ \citenamefont
  {Gallagher}}]{PhysRevLett.58.1324}%
  \BibitemOpen
  \bibfield  {author} {\bibinfo {author} {\bibfnamefont {R.~C.}\ \bibnamefont
  {Stoneman}}, \bibinfo {author} {\bibfnamefont {M.~D.}\ \bibnamefont {Adams}},
  \ and\ \bibinfo {author} {\bibfnamefont {T.~F.}\ \bibnamefont {Gallagher}},\
  }\href {\doibase 10.1103/PhysRevLett.58.1324} {\bibfield  {journal} {\bibinfo
   {journal} {Phys. Rev. Lett.}\ }\textbf {\bibinfo {volume} {58}},\ \bibinfo
  {pages} {1324} (\bibinfo {year} {1987})}\BibitemShut {NoStop}%
\bibitem [{\citenamefont {LINNEWEBER}\ \emph {et~al.}(2012)\citenamefont
  {LINNEWEBER}, \citenamefont {STOLZE},\ and\ \citenamefont {UHRIG}}]{pla_39}%
  \BibitemOpen
  \bibfield  {author} {\bibinfo {author} {\bibfnamefont {T.}~\bibnamefont
  {LINNEWEBER}}, \bibinfo {author} {\bibfnamefont {J.}~\bibnamefont {STOLZE}},
  \ and\ \bibinfo {author} {\bibfnamefont {G.~S.}\ \bibnamefont {UHRIG}},\
  }\href {\doibase 10.1142/S0219749912500293} {\bibfield  {journal} {\bibinfo
  {journal} {International Journal of Quantum Information}\ }\textbf {\bibinfo
  {volume} {10}},\ \bibinfo {pages} {1250029} (\bibinfo {year}
  {2012})}\BibitemShut {NoStop}%
\bibitem [{\citenamefont {Lorenzo}\ \emph {et~al.}(2013)\citenamefont
  {Lorenzo}, \citenamefont {Apollaro}, \citenamefont {Sindona},\ and\
  \citenamefont {Plastina}}]{pla_40}%
  \BibitemOpen
  \bibfield  {author} {\bibinfo {author} {\bibfnamefont {S.}~\bibnamefont
  {Lorenzo}}, \bibinfo {author} {\bibfnamefont {T.~J.~G.}\ \bibnamefont
  {Apollaro}}, \bibinfo {author} {\bibfnamefont {A.}~\bibnamefont {Sindona}}, \
  and\ \bibinfo {author} {\bibfnamefont {F.}~\bibnamefont {Plastina}},\ }\href
  {\doibase 10.1103/PhysRevA.87.042313} {\bibfield  {journal} {\bibinfo
  {journal} {Phys. Rev. A}\ }\textbf {\bibinfo {volume} {87}},\ \bibinfo
  {pages} {042313} (\bibinfo {year} {2013})}\BibitemShut {NoStop}%
\bibitem [{\citenamefont {Lorenzo}\ \emph {et~al.}(2015)\citenamefont
  {Lorenzo}, \citenamefont {Apollaro}, \citenamefont {Paganelli}, \citenamefont
  {Palma},\ and\ \citenamefont {Plastina}}]{pla_41}%
  \BibitemOpen
  \bibfield  {author} {\bibinfo {author} {\bibfnamefont {S.}~\bibnamefont
  {Lorenzo}}, \bibinfo {author} {\bibfnamefont {T.~J.~G.}\ \bibnamefont
  {Apollaro}}, \bibinfo {author} {\bibfnamefont {S.}~\bibnamefont {Paganelli}},
  \bibinfo {author} {\bibfnamefont {G.~M.}\ \bibnamefont {Palma}}, \ and\
  \bibinfo {author} {\bibfnamefont {F.}~\bibnamefont {Plastina}},\ }\href
  {\doibase 10.1103/PhysRevA.91.042321} {\bibfield  {journal} {\bibinfo
  {journal} {Phys. Rev. A}\ }\textbf {\bibinfo {volume} {91}},\ \bibinfo
  {pages} {042321} (\bibinfo {year} {2015})}\BibitemShut {NoStop}%
\bibitem [{\citenamefont {Almeida}\ \emph {et~al.}(2016)\citenamefont
  {Almeida}, \citenamefont {Ciccarello}, \citenamefont {Apollaro},\ and\
  \citenamefont {Souza}}]{pla_42}%
  \BibitemOpen
  \bibfield  {author} {\bibinfo {author} {\bibfnamefont {G.~M.~A.}\
  \bibnamefont {Almeida}}, \bibinfo {author} {\bibfnamefont {F.}~\bibnamefont
  {Ciccarello}}, \bibinfo {author} {\bibfnamefont {T.~J.~G.}\ \bibnamefont
  {Apollaro}}, \ and\ \bibinfo {author} {\bibfnamefont {A.~M.~C.}\ \bibnamefont
  {Souza}},\ }\href {\doibase 10.1103/PhysRevA.93.032310} {\bibfield  {journal}
  {\bibinfo  {journal} {Phys. Rev. A}\ }\textbf {\bibinfo {volume} {93}},\
  \bibinfo {pages} {032310} (\bibinfo {year} {2016})}\BibitemShut {NoStop}%
\bibitem [{\citenamefont {Flannery}\ \emph {et~al.}(2005)\citenamefont
  {Flannery}, \citenamefont {Vrinceanu},\ and\ \citenamefont
  {Ostrovsky}}]{thadwalkerflannery}%
  \BibitemOpen
  \bibfield  {author} {\bibinfo {author} {\bibfnamefont {M.~R.}\ \bibnamefont
  {Flannery}}, \bibinfo {author} {\bibfnamefont {D.}~\bibnamefont {Vrinceanu}},
  \ and\ \bibinfo {author} {\bibfnamefont {V.~N.}\ \bibnamefont {Ostrovsky}},\
  }\href@noop {} {\bibfield  {journal} {\bibinfo  {journal} {J. Phys. B. At.
  Mol. Opt. Phys.}\ }\textbf {\bibinfo {volume} {38}},\ \bibinfo {pages} {S279}
  (\bibinfo {year} {2005})}\BibitemShut {NoStop}%
\bibitem [{\citenamefont {Reinhard}\ \emph {et~al.}(2007)\citenamefont
  {Reinhard}, \citenamefont {Liebisch}, \citenamefont {Knuffman},\ and\
  \citenamefont {Raithel}}]{thadwalkerreihard}%
  \BibitemOpen
  \bibfield  {author} {\bibinfo {author} {\bibfnamefont {A.}~\bibnamefont
  {Reinhard}}, \bibinfo {author} {\bibfnamefont {T.~C.}\ \bibnamefont
  {Liebisch}}, \bibinfo {author} {\bibfnamefont {B.}~\bibnamefont {Knuffman}},
  \ and\ \bibinfo {author} {\bibfnamefont {G.}~\bibnamefont {Raithel}},\ }\href
  {\doibase 10.1103/PhysRevA.75.032712} {\bibfield  {journal} {\bibinfo
  {journal} {Phys. Rev. A}\ }\textbf {\bibinfo {volume} {75}},\ \bibinfo
  {pages} {032712} (\bibinfo {year} {2007})}\BibitemShut {NoStop}%
\bibitem [{\citenamefont {Singer}\ \emph {et~al.}(2005)\citenamefont {Singer},
  \citenamefont {Stanojevic}, \citenamefont {Weidemüller},\ and\ \citenamefont
  {Cote}}]{thadwalkersinger}%
  \BibitemOpen
  \bibfield  {author} {\bibinfo {author} {\bibfnamefont {K.}~\bibnamefont
  {Singer}}, \bibinfo {author} {\bibfnamefont {J.}~\bibnamefont {Stanojevic}},
  \bibinfo {author} {\bibfnamefont {M.}~\bibnamefont {Weidemüller}}, \ and\
  \bibinfo {author} {\bibfnamefont {R.}~\bibnamefont {Cote}},\ }\href@noop {}
  {\bibfield  {journal} {\bibinfo  {journal} {J. Phys. B. At. Mol. Opt. Phys.}\
  }\textbf {\bibinfo {volume} {38}},\ \bibinfo {pages} {S295} (\bibinfo {year}
  {2005})}\BibitemShut {NoStop}%
\bibitem [{\citenamefont {F¨orster}(1996)}]{for96}%
  \BibitemOpen
  \bibfield  {author} {\bibinfo {author} {\bibfnamefont {T.}~\bibnamefont
  {F¨orster}},\ }\href@noop {} {\emph {\bibinfo {title} {Modern Quantum
  Chemistry, edited by O. Sinanoglu}}}\ (\bibinfo  {publisher} {Academic Press,
  New York},\ \bibinfo {year} {1996})\BibitemShut {NoStop}%
\bibitem [{\citenamefont {Westermann}\ \emph {et~al.}(2006)\citenamefont
  {Westermann}, \citenamefont {Amthor}, \citenamefont {de~Oliveira},
  \citenamefont {Deiglmayr}, \citenamefont {Reetz-Lamour},\ and\ \citenamefont
  {Weidemüller}}]{epjd}%
  \BibitemOpen
  \bibfield  {author} {\bibinfo {author} {\bibfnamefont {S.}~\bibnamefont
  {Westermann}}, \bibinfo {author} {\bibfnamefont {T.}~\bibnamefont {Amthor}},
  \bibinfo {author} {\bibfnamefont {A.}~\bibnamefont {de~Oliveira}}, \bibinfo
  {author} {\bibfnamefont {J.}~\bibnamefont {Deiglmayr}}, \bibinfo {author}
  {\bibfnamefont {M.}~\bibnamefont {Reetz-Lamour}}, \ and\ \bibinfo {author}
  {\bibfnamefont {M.}~\bibnamefont {Weidemüller}},\ }\href@noop {} {\bibfield
  {journal} {\bibinfo  {journal} {Eur. Phys. J. D}\ }\textbf {\bibinfo {volume}
  {40}},\ \bibinfo {pages} {37} (\bibinfo {year} {2006})}\BibitemShut {NoStop}%
\bibitem [{\citenamefont {Comparat}\ and\ \citenamefont
  {Pillet}(2010)}]{Comparat}%
  \BibitemOpen
  \bibfield  {author} {\bibinfo {author} {\bibfnamefont {D.}~\bibnamefont
  {Comparat}}\ and\ \bibinfo {author} {\bibfnamefont {P.}~\bibnamefont
  {Pillet}},\ }\href {\doibase 10.1364/JOSAB.27.00A208} {\bibfield  {journal}
  {\bibinfo  {journal} {J. Opt. Soc. Am. B}\ }\textbf {\bibinfo {volume}
  {27}},\ \bibinfo {pages} {A208} (\bibinfo {year} {2010})}\BibitemShut
  {NoStop}%
\bibitem [{\citenamefont {Jackson}(1999)}]{jd}%
  \BibitemOpen
  \bibfield  {author} {\bibinfo {author} {\bibfnamefont {J.~D.}\ \bibnamefont
  {Jackson}},\ }\href@noop {} {\emph {\bibinfo {title} {Classical
  Electrodynamics}}},\ \bibinfo {edition} {3rd}\ ed.\ (\bibinfo  {publisher}
  {John Wiley \& Sons Inc.},\ \bibinfo {address} {New York},\ \bibinfo {year}
  {1999})\ pp.\ \bibinfo {pages} {xxii+808}\BibitemShut {NoStop}%
\bibitem [{\citenamefont {Anderson}\ \emph {et~al.}(1998)\citenamefont
  {Anderson}, \citenamefont {Veale},\ and\ \citenamefont
  {Gallagher}}]{pair_int5}%
  \BibitemOpen
  \bibfield  {author} {\bibinfo {author} {\bibfnamefont {W.~R.}\ \bibnamefont
  {Anderson}}, \bibinfo {author} {\bibfnamefont {J.~R.}\ \bibnamefont {Veale}},
  \ and\ \bibinfo {author} {\bibfnamefont {T.~F.}\ \bibnamefont {Gallagher}},\
  }\href {\doibase 10.1103/PhysRevLett.80.249} {\bibfield  {journal} {\bibinfo
  {journal} {Phys. Rev. Lett.}\ }\textbf {\bibinfo {volume} {80}},\ \bibinfo
  {pages} {249} (\bibinfo {year} {1998})}\BibitemShut {NoStop}%
\bibitem [{\citenamefont {G{\"u}nter}\ \emph {et~al.}(2013)\citenamefont
  {G{\"u}nter}, \citenamefont {Schempp}, \citenamefont {Robert-de
  Saint-Vincent}, \citenamefont {Gavryusev}, \citenamefont {Helmrich},
  \citenamefont {Hofmann}, \citenamefont {Whitlock},\ and\ \citenamefont
  {Weidem{\"u}ller}}]{pair_int37}%
  \BibitemOpen
  \bibfield  {author} {\bibinfo {author} {\bibfnamefont {G.}~\bibnamefont
  {G{\"u}nter}}, \bibinfo {author} {\bibfnamefont {H.}~\bibnamefont {Schempp}},
  \bibinfo {author} {\bibfnamefont {M.}~\bibnamefont {Robert-de
  Saint-Vincent}}, \bibinfo {author} {\bibfnamefont {V.}~\bibnamefont
  {Gavryusev}}, \bibinfo {author} {\bibfnamefont {S.}~\bibnamefont {Helmrich}},
  \bibinfo {author} {\bibfnamefont {C.~S.}\ \bibnamefont {Hofmann}}, \bibinfo
  {author} {\bibfnamefont {S.}~\bibnamefont {Whitlock}}, \ and\ \bibinfo
  {author} {\bibfnamefont {M.}~\bibnamefont {Weidem{\"u}ller}},\ }\href
  {\doibase 10.1126/science.1244843} {\bibfield  {journal} {\bibinfo  {journal}
  {Science}\ }\textbf {\bibinfo {volume} {342}},\ \bibinfo {pages} {954}
  (\bibinfo {year} {2013})},\ \Eprint
  {http://arxiv.org/abs/http://science.sciencemag.org/content/342/6161/954.full.pdf}
  {http://science.sciencemag.org/content/342/6161/954.full.pdf} \BibitemShut
  {NoStop}%
\bibitem [{\citenamefont {Barredo}\ \emph {et~al.}(2014)\citenamefont
  {Barredo}, \citenamefont {Ravets}, \citenamefont {Labuhn}, \citenamefont
  {B\'eguin}, \citenamefont {Vernier}, \citenamefont {Nogrette}, \citenamefont
  {Lahaye},\ and\ \citenamefont {Browaeys}}]{pair_int38}%
  \BibitemOpen
  \bibfield  {author} {\bibinfo {author} {\bibfnamefont {D.}~\bibnamefont
  {Barredo}}, \bibinfo {author} {\bibfnamefont {S.}~\bibnamefont {Ravets}},
  \bibinfo {author} {\bibfnamefont {H.}~\bibnamefont {Labuhn}}, \bibinfo
  {author} {\bibfnamefont {L.}~\bibnamefont {B\'eguin}}, \bibinfo {author}
  {\bibfnamefont {A.}~\bibnamefont {Vernier}}, \bibinfo {author} {\bibfnamefont
  {F.}~\bibnamefont {Nogrette}}, \bibinfo {author} {\bibfnamefont
  {T.}~\bibnamefont {Lahaye}}, \ and\ \bibinfo {author} {\bibfnamefont
  {A.}~\bibnamefont {Browaeys}},\ }\href {\doibase
  10.1103/PhysRevLett.112.183002} {\bibfield  {journal} {\bibinfo  {journal}
  {Phys. Rev. Lett.}\ }\textbf {\bibinfo {volume} {112}},\ \bibinfo {pages}
  {183002} (\bibinfo {year} {2014})}\BibitemShut {NoStop}%
\bibitem [{\citenamefont {Barredo}\ \emph {et~al.}(2015)\citenamefont
  {Barredo}, \citenamefont {Labuhn}, \citenamefont {Ravets}, \citenamefont
  {Lahaye}, \citenamefont {Browaeys},\ and\ \citenamefont {Adams}}]{danielb}%
  \BibitemOpen
  \bibfield  {author} {\bibinfo {author} {\bibfnamefont {D.}~\bibnamefont
  {Barredo}}, \bibinfo {author} {\bibfnamefont {H.}~\bibnamefont {Labuhn}},
  \bibinfo {author} {\bibfnamefont {S.}~\bibnamefont {Ravets}}, \bibinfo
  {author} {\bibfnamefont {T.}~\bibnamefont {Lahaye}}, \bibinfo {author}
  {\bibfnamefont {A.}~\bibnamefont {Browaeys}}, \ and\ \bibinfo {author}
  {\bibfnamefont {C.~S.}\ \bibnamefont {Adams}},\ }\href {\doibase
  10.1103/PhysRevLett.114.113002} {\bibfield  {journal} {\bibinfo  {journal}
  {Phys. Rev. Lett.}\ }\textbf {\bibinfo {volume} {114}},\ \bibinfo {pages}
  {113002} (\bibinfo {year} {2015})}\BibitemShut {NoStop}%
\bibitem [{\citenamefont {Maineult}\ \emph {et~al.}(2016)\citenamefont
  {Maineult}, \citenamefont {Pelle}, \citenamefont {Faoro}, \citenamefont
  {Arimondo}, \citenamefont {Pillet},\ and\ \citenamefont
  {Cheinet}}]{pair_int40}%
  \BibitemOpen
  \bibfield  {author} {\bibinfo {author} {\bibfnamefont {W.}~\bibnamefont
  {Maineult}}, \bibinfo {author} {\bibfnamefont {B.}~\bibnamefont {Pelle}},
  \bibinfo {author} {\bibfnamefont {R.}~\bibnamefont {Faoro}}, \bibinfo
  {author} {\bibfnamefont {E.}~\bibnamefont {Arimondo}}, \bibinfo {author}
  {\bibfnamefont {P.}~\bibnamefont {Pillet}}, \ and\ \bibinfo {author}
  {\bibfnamefont {P.}~\bibnamefont {Cheinet}},\ }\href
  {http://stacks.iop.org/0953-4075/49/i=21/a=214001} {\bibfield  {journal}
  {\bibinfo  {journal} {J. Phys. B. At. Mol. Opt. Phys.}\ }\textbf {\bibinfo
  {volume} {49}},\ \bibinfo {pages} {214001} (\bibinfo {year}
  {2016})}\BibitemShut {NoStop}%
\bibitem [{\citenamefont {Scholes}(2003)}]{sho3}%
  \BibitemOpen
  \bibfield  {author} {\bibinfo {author} {\bibfnamefont {G.~D.}\ \bibnamefont
  {Scholes}},\ }\href {\doibase 10.1146/annurev.physchem.54.011002.103746}
  {\bibfield  {journal} {\bibinfo  {journal} {Annu. Rev. Phys. Chem.}\ }\textbf
  {\bibinfo {volume} {54}},\ \bibinfo {pages} {57} (\bibinfo {year}
  {2003})}\BibitemShut {NoStop}%
\bibitem [{\citenamefont {Ritz}\ \emph {et~al.}(2002)\citenamefont {Ritz},
  \citenamefont {Damjanović},\ and\ \citenamefont {Schulten}}]{rit2}%
  \BibitemOpen
  \bibfield  {author} {\bibinfo {author} {\bibfnamefont {T.}~\bibnamefont
  {Ritz}}, \bibinfo {author} {\bibfnamefont {A.}~\bibnamefont {Damjanović}}, \
  and\ \bibinfo {author} {\bibfnamefont {K.}~\bibnamefont {Schulten}},\ }\href
  {\doibase 10.1002/1439-7641(20020315)3:3<243::AID-CPHC243>3.0.CO;2-Y}
  {\bibfield  {journal} {\bibinfo  {journal} {Chem. Phys. Chem.}\ }\textbf
  {\bibinfo {volume} {3}},\ \bibinfo {pages} {243} (\bibinfo {year}
  {2002})}\BibitemShut {NoStop}%
\bibitem [{\citenamefont {Schau{\ss}}\ \emph {et~al.}(2015)\citenamefont
  {Schau{\ss}}, \citenamefont {Zeiher}, \citenamefont {Fukuhara}, \citenamefont
  {Hild}, \citenamefont {Cheneau}, \citenamefont {Macr{\`\i}}, \citenamefont
  {Pohl}, \citenamefont {Bloch},\ and\ \citenamefont {Gross}}]{pair_int41}%
  \BibitemOpen
  \bibfield  {author} {\bibinfo {author} {\bibfnamefont {P.}~\bibnamefont
  {Schau{\ss}}}, \bibinfo {author} {\bibfnamefont {J.}~\bibnamefont {Zeiher}},
  \bibinfo {author} {\bibfnamefont {T.}~\bibnamefont {Fukuhara}}, \bibinfo
  {author} {\bibfnamefont {S.}~\bibnamefont {Hild}}, \bibinfo {author}
  {\bibfnamefont {M.}~\bibnamefont {Cheneau}}, \bibinfo {author} {\bibfnamefont
  {T.}~\bibnamefont {Macr{\`\i}}}, \bibinfo {author} {\bibfnamefont
  {T.}~\bibnamefont {Pohl}}, \bibinfo {author} {\bibfnamefont {I.}~\bibnamefont
  {Bloch}}, \ and\ \bibinfo {author} {\bibfnamefont {C.}~\bibnamefont
  {Gross}},\ }\href {\doibase 10.1126/science.1258351} {\bibfield  {journal}
  {\bibinfo  {journal} {Science}\ }\textbf {\bibinfo {volume} {347}},\ \bibinfo
  {pages} {1455} (\bibinfo {year} {2015})},\ \Eprint
  {http://arxiv.org/abs/http://science.sciencemag.org/content/347/6229/1455.full.pdf}
  {http://science.sciencemag.org/content/347/6229/1455.full.pdf} \BibitemShut
  {NoStop}%
\bibitem [{\citenamefont {Hito}\ \emph {et~al.}(2018)\citenamefont {Hito},
  \citenamefont {Silva},\ and\ \citenamefont {de~Magalhães}}]{pla}%
  \BibitemOpen
  \bibfield  {author} {\bibinfo {author} {\bibfnamefont {C.~B.}\ \bibnamefont
  {Hito}}, \bibinfo {author} {\bibfnamefont {M.}~\bibnamefont {Silva}}, \ and\
  \bibinfo {author} {\bibfnamefont {A.~B.}\ \bibnamefont {de~Magalhães}},\
  }\href {\doibase https://doi.org/10.1016/j.physleta.2018.01.033} {\bibfield
  {journal} {\bibinfo  {journal} {Physics Letters A}\ }\textbf {\bibinfo
  {volume} {382}},\ \bibinfo {pages} {894 } (\bibinfo {year}
  {2018})}\BibitemShut {NoStop}%
\bibitem [{\citenamefont {{Blum}}(2012)}]{bookblum}%
  \BibitemOpen
  \bibfield  {author} {\bibinfo {author} {\bibfnamefont {K.}~\bibnamefont
  {{Blum}}},\ }\href {\doibase 10.1007/978-3-642-20561-3} {\emph {\bibinfo
  {title} {{Density matrix theory and applications}}}},\ \bibinfo {edition}
  {3rd}\ ed.\ (\bibinfo  {publisher} {Berlin: Springer},\ \bibinfo {year}
  {2012})\ pp.\ \bibinfo {pages} {xviii + 343}\BibitemShut {NoStop}%
\bibitem [{\citenamefont {Gounand}(1979)}]{gounand}%
  \BibitemOpen
  \bibfield  {author} {\bibinfo {author} {\bibfnamefont {F.}~\bibnamefont
  {Gounand}},\ }\href {\doibase 10.1051/jphys:01979004005045700} {\bibfield
  {journal} {\bibinfo  {journal} {J. Phys. France}\ }\textbf {\bibinfo {volume}
  {40}},\ \bibinfo {pages} {457} (\bibinfo {year} {1979})}\BibitemShut
  {NoStop}%
\bibitem [{\citenamefont {Branden}\ \emph {et~al.}(2010)\citenamefont
  {Branden}, \citenamefont {Juhasz}, \citenamefont {Mahlokozera}, \citenamefont
  {Vesa}, \citenamefont {Wilson}, \citenamefont {Zheng}, \citenamefont
  {Kortyna},\ and\ \citenamefont {Tate}}]{branden}%
  \BibitemOpen
  \bibfield  {author} {\bibinfo {author} {\bibfnamefont {D.~B.}\ \bibnamefont
  {Branden}}, \bibinfo {author} {\bibfnamefont {T.}~\bibnamefont {Juhasz}},
  \bibinfo {author} {\bibfnamefont {T.}~\bibnamefont {Mahlokozera}}, \bibinfo
  {author} {\bibfnamefont {C.}~\bibnamefont {Vesa}}, \bibinfo {author}
  {\bibfnamefont {R.~O.}\ \bibnamefont {Wilson}}, \bibinfo {author}
  {\bibfnamefont {M.}~\bibnamefont {Zheng}}, \bibinfo {author} {\bibfnamefont
  {A.}~\bibnamefont {Kortyna}}, \ and\ \bibinfo {author} {\bibfnamefont
  {D.~A.}\ \bibnamefont {Tate}},\ }\href
  {http://stacks.iop.org/0953-4075/43/i=1/a=015002} {\bibfield  {journal}
  {\bibinfo  {journal} {J. Phys. B. At. Mol. Opt. Phys.}\ }\textbf {\bibinfo
  {volume} {43}},\ \bibinfo {pages} {015002} (\bibinfo {year}
  {2010})}\BibitemShut {NoStop}%
\end{thebibliography}
\end{document}